\documentclass[prd, preprint, longbibliography, 12pt]{revtex4-1}

\usepackage{amsmath}
\usepackage{amssymb}
\usepackage{setspace}
\usepackage{graphicx}
\usepackage{natbib}
\usepackage{float}
\usepackage{tensor}
\usepackage[utf8]{inputenc}
\usepackage{amsfonts}
\usepackage{braket}
\usepackage{esint}
\usepackage{breqn}
\usepackage{IEEEtrantools}
\usepackage{hyperref}

\begin{document}
 
 %

\begin{center}
{ \large \bf Black Hole Entropy from Trace Dynamics and Non-commutative Geometry}



\vskip 0.2 in

{\large{\bf Palemkota Maithresh$^{a1}$ and Tejinder P.  Singh$^{b2}$ }}

\medskip

{\it $^a${UM-DAE Centre for Excellence in Basic Sciences, Mumbai, 400098, India}\\

{\it $^b$Tata Institute of Fundamental Research,}
{\it Homi Bhabha Road, Mumbai 400005, India}\\
\bigskip
 {$^{1}$\tt p.maithresh@cbs.ac.in}, \; {$^2$\tt tpsingh@tifr.res.in}}

\end{center}

\bigskip

\centerline{\bf ABSTRACT}
\noindent  Spontaneous localisation is a falsifiable, phenomenological, mechanism for explaining the absence of macroscopic position superpositions, currently being tested for in the laboratory. The theory of trace dynamics provides a possible theoretical origin for spontaneous localisation. We have recently proposed how to employ non-commutative geometry to include gravity in trace dynamics, and suggested the emergence of classical space-time geometry via spontaneous localisation. 
In our theory, which we call non-commutative matter gravity,  a black hole arises from the spontaneous localisation of an entangled state of a large number of `atoms of space-time-matter [STM]'. Prior to localisation, the non-commutative curvature of an STM atom is described  by the spectral action of non-commutative geometry. By using the techniques of statistical thermodynamics from trace dynamics, we show that the gravitational entropy of a Schwarzschild black hole results from the microstates of the entangled STM atoms and is given (subject to certain assumptions) by the classical   Euclidean gravitational action. This action,  in turn, equals the Bekenstein-Hawking entropy (Area/$4{L_P}^2$) of the black hole. We argue that spontaneous localisation is related to black-hole evaporation through the fluctuation-dissipation theorem.

\bigskip

\bigskip

\section{Introduction}
Spontaneous localisation is a falsifiable mechanism for explaining the observed absence of position superposition in macroscopic objects \cite{Ghirardi:86, Ghirardi2:90, Pearle:89,  Bassi:03, RMP:2012}. It involves a small modification of quantum theory. Quantum mechanics [i.e. the Schrodinger equation] says that a position superposition (say superposition of two position eigenstates of an electron) once created, lasts forever. The theory of spontaneous localisation disagrees slightly, and proposes that the superposition of two position states does not last forever, but for a very large time $T$ (this being the mean life-time). After a time $T$, the superposition spontaneously and randomly decays (according to a Poisson process in time)  to one of the superposed states.   For definiteness., it is proposed that for a nucleon this new constant of nature, $T$, is assumed to take the value $10^{17}$ s. This is enough to explain the absence of macroscopic superpositions. This is because for an entangled state of $N$ nucleons, the superposition lifetime scales down to $T/N$, because every nucleon's localisation event is independent of the other, and localisation of any one out of the $N$ entangled nucleons localises the entire system of $N$ particles. Since for a macroscopic object $N$ is enormous, $T/N$ is very small. Hence, microscopic superpositions last for very long times, but not forever. On the other hand, macroscopic superpositions are not completely absent, but extremely short-lived. In this way, through a slight modification of quantum mechanics, the theory of spontaneous localisation explains the quantum-to-classical transition dynamically. Experiments are tightening the experimental bound on $T$, the current bound being that 
$T>10^{8}$ s \cite{Mauro2019}. If experiments push the bound beyond $10^{17}$ s, the theory would be ruled out, as then macro-superpositions will be long lived, contrary to experience.

While there is no confirmed direct experimental  evidence for quantising gravity, it can be argued that there is a strong theoretical motivation. We have argued earlier that there ought to exist an equivalent reformulation of quantum (field) theory which does not refer to classical space-time \cite{Singh:2012,Singh:2006}. Such a reformulation is inevitably also a quantum theory of gravity. Moreover, we argue that a quantum theory of gravity must dynamically explain the absence of superposition of space-time geometries in the classical limit where general relativity is recovered. Spontaneous localisation of space-time superpositions is a plausible mechanism for achieving this. Hence there appears to be a deep connection between the absence of position superposition in macroscopic objects, and the emergence of classical space-time from quantum gravity. This aspect has been developed in our work described below \cite{maithresh2019}.
 
 The theory of spontaneous localisation is phenomenological and ad hoc, having been proposed with the express purpose of explaining the quantum-to-classical transition. It introduces $T$ as a new constant of nature, and moreover the theory is non-relativistic. Attempts to include spontaneous localisation in relativistic quantum field theory face challenges. It is desirable to understand the theoretical origins of the proposal - what is the source of the stochastic noise which causes collapse to occur spontaneously, and what is the spectrum of this noise? The theory of trace dynamics developed by Adler and collaborators \cite{Adler:04, Adler:94, RMP:2012, Pearle:2005}. provides a framework for understanding spontaneous collapse. Trace dynamics is a matrix dynamics of Grassmann valued matrices, possessing a global unitary invariance. In this dynamics, classical point particles (as well as classical fields) are raised to the status of operators/matrices. The theory operates at the Planck scale; gravity is not included however. One asks as to what trace dynamics looks like at energy scales much lower than Planck scale, i.e. when coarse-grained over time intervals much larger than Planck time. The methods of statistical thermodynamics are employed to show that the emergent dynamics at statistical equilibrium is relativistic quantum (field) theory. Moreover, under certain circumstances, statistical fluctuations around equilibrium become significant - these are then responsible for spontaneous localisation, leading to the emergence of classical behaviour in macroscopic objects.
 
 Trace dynamics does not include gravity. In our recent work \cite{maithresh2019}, we have proposed as to how the principles of non-commutative geometry (Connes' program) \cite{Connes2000} can be employed to incorporate gravity into trace dynamics. In analogy with trace dynamics, now space-time points are raised to the status of operators/matrices, and this generalised trace dynamics is also assumed to operate at the Planck scale. The Dirac operator relates to the curvature of a Riemannian manifold, and this operator also serves to define curvature in a non-commutative geometry. Time evolution is described by a parameter unique to 
 non-commutative geometry. The Dirac operator is utilised to construct a trace dynamics Lagrangian and action principle for gravity. Matter fermions are introduced by proposing the concept of `atoms' of space-time-matter that evolve in Hilbert space. This is the theory of non-commutative matter-gravity operating at the Planck scale. It is the generalisation of trace dynamics to include gravity. The statistical thermodynamics of this generalised trace dynamics yields a theory of quantum gravity at equilibrium, at energies below Planck scale, after the underlying matrix dynamics has been coarse-grained over time intervals larger than Planck time. The implication is that quantum gravity is not a Planck scale phenomenon: rather, quantum gravity comes into play whenever no background space-time is available to describe quantum dynamics and associated gravity. For instance, we can ask how to describe the gravitation of an electron when it is in the superposed state in the double slit experiment? This is not Planck scale physics; yet the gravitation here cannot be described classically. Our emergent quantum theory explains how to describe gravitation under such circumstances. This quantum gravity is also the sought for description of quantum (field) theory without classical time. 
 
 The underlying matrix dynamics including gravity is in general non-unitary; even though the emergent quantum gravity at equilibrium is a unitary theory. The underlying theory admits a tiny anti-self-adjoint correction to the Hamiltonian of the theory.This tiny correction becomes amplified to a large effect if a large number of STM atoms get entangled with each other (in the same sense as in quantum entanglement). In this case, sub-Planckian fluctuations become important, leading to a modification of the equilibrium quantum gravity, and resulting in the spontaneous localisation of the (matter) fermionic part of the entangled STM atoms. This leads to the emergence of classical space-time and its Riemannian geometry, sourced by relativistic point-particle matter sources.
 
 In our theory, non-gravitational interactions and bosonic matter fields have not been included yet. Thus it is a theory only of Dirac fermions coupled to their self-gravity. In the classical limit, spontaneous localisation of matter fermions necessarily results in the formation of black holes. The entropy of such a black hole can be calculated from the microstates which are compatible with the emergence of a black hole with given macroscopic parameters. In the present paper we use our formalism to prove that the computed black hole entropy of a Schwarzschild black hole equals the Bekenstein-Hawking entropy: one-fourth the black hole area, in Planck units.
 
 The paper is divided as follows. In Section II we review the essential physics of trace dynamics. In Sections III and IV, we review our recent work on including gravity in trace dynamics, and how classical general relativity emerges after spontaneous localisation. These results are included here so as to make the present paper more self-contained and readable. Section V presents our new results on calculation of black hole entropy in our approach.

\section{Spontaneous Localisation from Trace Dynamics}
Trace dynamics [TD] derives quantum (field) theory and spontaneous localisation from an underlying (pre-quantum) matrix dynamics. It is the dynamics of matrix models obeying a global unitary invariance, operating at the Planck scale; [however, space-time is assumed to be Minkowski space-time, and gravity is not included in the theory]. Suppose we take classical dynamics [either Newtonian mechanics or special relativity] as the starting point, and instead of describing a material point particle by a real number, we describe it by a matrix (equivalently, operator). This is the essence of trace dynamics: for a particle $q$, now described by a matrix ${\bf q}$, the action is transformed as in this example:
\begin{equation}
S = \int dt \ [\dot{q}^2 - q^2] \quad  \longrightarrow \int dt \ Tr\  [\dot{\bf q}^2 - {\bf q}^2]
\end{equation}
After replacing the configuration variable $q$ by a matrix, the scalar Lagrangian is constructed by taking a matrix trace of the operator polynomial, and then the scalar action is constructed as usual,  by integrating the Lagrangian over time. A general trace Lagrangian ${\bf L}$ would be a function of the various configuration variables ${\bf q}_i$ and their time derivatives $\dot{\bf q}_i$, and would be made from the trace of an operator polynomial ${\cal L}$. This construction can be extended to field theory by raising the field value at each space-time point to a matrix, constructing an operator polynomial, taking its trace to form a Lagrangian density, and integrating over four-volume to get the action. 

Lagrange equations of motion are obtained by varying the action with respect to the operator ${\bf q}_i$. In order to vary the trace Lagrangian with respect to an operator, the notion of a trace derivative is introduced. The derivative of the trace Lagrangian ${\bf L}$  with respect to an operator $\cal O$ in the polynomial ${\cal L}$ is defined as
\begin{equation}
\delta {\bf L }= Tr \frac{\delta{\bf L}}{\delta\cal{O}}\delta\cal{O}
\end{equation}
This so-called trace derivative is obtained by varying ${\bf L}$ with respect to ${\cal O}$ and then cyclically permuting ${\cal O}$ inside the trace, so that $\delta\cal O$ sits to the right of the polynomial $\cal{L}$. 

It is assumed that the matrix elements are complex valued Grassmann numbers, which can be further sub-divided into even grade and odd grade Grassmann numbers. Any Grassmann matrix can be split into a sum of two matrices: the bosonic part (made of even grade elements) and the fermionic part (made of odd grade elements). Bosonic (fermionic) matrices describe bosonic (fermionic) fields, as in conventional quantum field theory. Thus, in trace dynamics there are bosonic degrees of freedom ${\bf q}_B$ and fermionic degrees of freedom ${\bf q}_F$. 

The Lagrange equations
\begin{equation}
\frac {d\;}{dt}\Big ( \frac{\delta {\bf L}}{\delta \dot {\bf q}_i}\Big ) - \Big ( \frac {\delta {\bf L}}{\delta {\bf q}_i}\Big ) = 0
\end{equation} 
are used to obtain the operator equations of motion,  and they also define the canonical momenta. The configuration variables and the momenta do not commute amongst each other, and the commutation relations are arbitrary. This is what makes trace dynamics different from both classical dynamics as well as from quantum theory. Apart from the trace Hamiltonian, 
\begin{equation}
{\bf H}=\sum_i \mathrm{Tr}[p_{Fi}\dot{q}_{Fi}]+\sum_i \mathrm{Tr}[p_{Bi}\dot{q}_{Bi}]-\mathrm{Tr}\; \mathcal{L}
\end{equation}
there is another conserved charge of great significance, the Adler-Millard charge, denoted $\tilde{C}$. This charge is a consequence of a global unitary invariance of the trace Lagrangian and the trace Hamiltonian. It is given by
\begin{equation}
    \tilde{C} = \sum_{r\in B}[q_r,p_r] -\sum_{r\in F} \{q_r,p_r \} 
    \label{amc}
\end{equation}
[We henceforth drop the bold notation from the canonical variables, it being understood that we deal with matrix/operator valued canonical variables]. The Adler-Millard charge is unique to matrix dynamics, and plays a central role in emergence of quantum theory from trace dynamics. If the trace Hamiltonian is self-adjoint, then the Adler-Millard charge can be shown to be anti-self-adjoint. Were the TD Hamiltonian to have an anti-self-adjoint component, this conserved charge picks up a 
self-adjoint component - this will be important for us when we incorporate gravity in TD.

The Hamilton's equations of motion are given as
\begin{align}
    \frac{\delta \textbf{H}}{\delta q_r} =-\dot{p}_r, \quad \quad \frac{\delta \textbf{H}}{\delta p_r} =\epsilon_r \dot{q}_r
\end{align}
where $\epsilon_r = 1(-1)$ when $q_r$ is bosonic(fermionic). 

The above dynamics is Lorentz invariant, and is  assumed to take place at the Planck energy scale. TD does not specify the form of the fundamental Lagrangian of nature, though we will choose a particular form below when we incorporate gravity into TD. Since the physical systems that we observe and experiment with, operate at energy scales much lower than Planck energy and are not probed over Planck times, we ask the following question: What is the averaged description of trace dynamics, if we coarse grain (smear)  the trace dynamics over time intervals much larger than Planck times? We might imagine that there are extremely rapid variations in the canonical variables over Planck time scales, but there is a smoothed out dynamics at lower energies, where these rapid variations have been smeared over. The methods of statistical mechanics are employed, treating the underlying dynamics as `microscopic' degrees of freedom, to show that the emergent 
coarse-grained dynamics  is relativistic quantum (field) theory.

One starts by constructing the matrix dynamics phase space, with (the real and imaginary parts of) each element $(q_{r})_{lm}$ of $q_r$ being a (pair of) independent degrees of freedom in the phase space, along with the matrix component (again real and imaginary part) $(p_i)_{im}$ of the corresponding momentum. We use the symbol $x$ to denote $q$ or $p$.  A measure $d\mu$ is defined in the phase space, as
\begin{equation}
(x_r)_{mn} = (x_r)^{0}_{mn} + i (x_r)^{1}_{mn}; \quad d\mu = \prod_A d\mu^{A}; \quad d\mu^{A}=\prod_{r,m,n} d(x_r)^{A}_{mn}
\end{equation}
where $A=0,1$ and the components $(x_r)^{A}_{mn}$ are real numbers. This measure is conserved during evolution, and obeys Liouville's theorem. Moreover, the measure is invariant under infinitesimal operator shifts $x_r\rightarrow x_r + \delta x_r$. 

A phase space density distribution function $\rho[\{x_r\}]$ is defined in the matrix element phase space, which determines the probability of finding the system point in some particular infinitesimal volume in phase space. A canonical ensemble is constructed for a sufficiently large number of identical systems, each of which start evolving from arbitrary initial conditions in the phase space. It is assumed that over time intervals much larger than Planck time, the accessible region of the phase space [i.e. the region consistent with a conserved trace Hamiltonian and a conserved Adler-Millard charge] is uniformly populated, and hence that the long time average [the coarse-grained dynamics] can be determined from the ensemble average at any one given time.  This equilibrium dynamics is determined by maximising the Boltzmann entropy
\begin{equation}
\frac{S_E}{k_B}=-\int d\mu \; \rho\; \ln\rho
\label{entro}
\end{equation}
subject to the constraints that the ensemble-averaged trace Hamiltonian $\langle {\bf H} \rangle_{AV}$ and the ensemble averaged Adler-Millard charge $\langle {\bf \tilde{C}}\rangle_{AV}$ are conserved. These two constraints are imposed by introducing the Lagrange multipliers $\overline\tau$ and $\tilde{\lambda}$ respectively, where $\overline\tau$ is a constant with dimensions of inverse mass, and $\overline\lambda$ an anti-self-adjoint  matrix with dimensions of inverse action.  

Thus the phase space density distribution $\rho$ depends, apart from dynamical variables, on $\tilde{C}, \tilde{\lambda}, {\bf H},\overline \tau$ and can be written as $\rho(\tilde{C}, \tilde{\lambda}, {\bf H},\overline \tau)$. It can be further shown that the dependence on $\tilde{C}$ and $\tilde{\lambda}$ is of the form $Tr(\tilde{\lambda}\tilde{C})$, so we write $\rho = \rho(Tr[\tilde{\lambda}\tilde{C}],\overline\tau,{\bf H})$. It can be shown, subject to the assumption that the ensemble does not favour any one state in the ensemble over the other, that the canonical ensemble average of the Adler-Millard charge takes the form
\begin{equation}
\langle \tilde{C} \rangle_{AV}  = i_{eff} \hbar; \qquad i_{eff} = i\ diag(1,-1,1,-1...,1,-1)
\end{equation}
where the real constant $\hbar$ is eventually identified with Planck's constant, subsequent to the emergence of quantum dynamics.

The equilibrium distribution is found by maximising the function $-\cal{F}$ where
\begin{equation}
{\cal F} = \int d\mu\ \rho \log \rho + \theta \int d\mu\ \rho + \int d\mu\ \rho Tr \tilde{\lambda}\tilde{C} +\overline \tau \int d\mu\ \rho {\bf H}
\end{equation}
and gives the result
\begin{align}
    \rho = Z^{-1} \exp{(-\text{Tr} \Tilde{\lambda} \Tilde{C} -\tilde\tau \textbf{H})} \label{rho} \\
    Z = \int d \mu \; \exp{(-\text{Tr} \Tilde{\lambda}\Tilde{C} -\tilde\tau \textbf{H})} \label{z}
\end{align}
The entropy at equilibrium is given by
\begin{equation}
   \frac{S_E}{k_B} =  \log{Z} - \text{Tr}\tilde{\lambda} \frac{\partial \log{Z}}{\partial \tilde{\lambda}} -\tilde \tau \frac{\partial \log{Z}}{\partial \tilde\tau}  \label{entropy}
\end{equation}

What is the mean dynamics obeyed by the variables $\langle x\rangle_{AV}$, averaged over the canonical ensemble, at energy scales below Planck scale? To answer this, one derives certain Ward identities, as is done for  functional integrals in quantum field theory, in analogy with the proof for the equipartition theorem in statistical mechanics. These identities are a consequence of the invariance of the phase space measure under constant shifts of the dynamical variables. Thus, in conventional statistical mechanics,  the equipartition theorem is a consequence of the vanishing of the integral of a total divergence:
\begin{equation}
0 = \int d\mu \ \frac{\partial [x_r \exp(-\ \beta H)]}{\partial x_s}
\end{equation}
In the statistical mechanics of trace dynamics, we have for a general operator $\cal{O}$, that its average over the canonical ensemble is unchanged when a dynamical variable is varied:
\begin{equation}
0 =  \int d\mu\ \delta_{x_r} (\rho{\cal O})
\end{equation}
One chooses ${\cal O}$ to be the operator $Tr\{\tilde{C}, i_{eff}\}W$ where $W$ is any bosonic polynomial function of the dynamical variables, and carries out the above variation, taking $\rho$ to be the equilibrium phase space density distribution function. Thus we have 
\begin{equation}
0 = \int d\mu \ \delta_{x_r}\ \left[\exp \left(-Tr \tilde{\lambda}\tilde{C} - \tilde\tau {\bf H} \right)\  Tr \{\tilde{C},i_{eff}\}W\right]
\end{equation}
A very important assumption is made, namely that $\tilde{\tau}$ is the Planck time scale, and that we are interested in the averaged dynamics over much larger time scales (equivalently much lower energies). Each dynamical variable $x_r$ is split into a `fast' varying part [which varies over Planck times] and a `slow' part which is constant over Planck times. Important conclusions then follow from the above Ward identity, by making different choices for $W$. When $W$ is chosen to be a dynamical variable $x_r$, standard quantum commutation relations for bosonic and fermionic degrees of freedom are shown to be obeyed by the averaged variables $\langle x_r\rangle_{AV}$. The constant $\hbar$ introduced above is identified with Planck's constant. If $W$ is identified with the operator polynomial $H$ whose trace is the trace Hamiltonian ${\bf H}$, the quantum Heisenberg equations of motion for the averaged dynamical variables are obtained. The underlying matrices of TD, within ensemble averages, obey properties analogous to quantum fields. The contact with quantum field theory is made as follows. There is a unique eigenvector $\psi_0$ whose corresponding eigenvalue is the lowest eigenvalue of $H$. This acts as the conventional vacuum state, and canonical ensemble averages are identified with Wightman functions in the emergent quantum field theory, for a given function S,
\begin{equation}
\psi^{\dagger}_0 \  \langle S \{x_r\}\rangle_{AV} \ \psi_0 = \langle vac | S\{X\}|vac\rangle
\end{equation}
where $X$ is a quantum field operator.
In this way, relativistic quantum
(field) theory is shown to be an emergent phenomenon, being the low energy equilibrium approximation in the statistical thermodynamics of an underlying matrix dynamics. Once the Heisenberg equations of motion are known, one can also transform to the Schrodinger picture in the standard manner.

Trace dynamics also provides a theoretical basis for the origin of the phenomenological theory of spontaneous localisation. As we have seen, quantum dynamics is a mean dynamics arising from averaging over Planck time scales, and neglecting the fast component in the variation of the dynamical variables. Under certain circumstances, the fast component can become significant, in which case its impact on the coarse-grained dynamics can be modelled as stochastic fluctuations around equilibrium. Particularly crucial is that these fluctuations can make an ant-self-adjoint stochastic contribution to the quantum theory Hamiltonian. This is possible because the underlying trace Hamiltonian can have a small anti-self-adjoint part at the Planck scale, which could get amplified by entanglement between a very large number of particles. Precisely such a situation arises when gravity is included in trace dynamics, as we will see below.

Adler considers such a possibility for fermions, in the non-relativistic approximation to quantum field theory, where the anti-Hermitean fluctuating correction to the Hamiltonian is modelled by adding a stochastic function ${\cal K}(t)$ 
\begin{equation}
i\hbar \frac{\partial \Psi}{\partial t} = H\Psi + i{\cal K}(t)\Psi
\end{equation}
This modified equation does not preserve norm of the state vector during evolution. If we insist on norm-preservation, and transform to a new state vector whose norm is preserved, the resulting evolution equation is non-linear. It also makes the evolution non-unitary; if we also demand that the non-linear evolution should not lead to superluminal signalling, the form of the evolution becomes just the same as in spontaneous localisation models. Thus trace dynamics can in principle explain the quantum-to-classical transition, by taking into account the potential role of statistical fluctuations around equilibrium. The theory provides a common origin for quantum theory, as well as for spontaneous localisation, starting from an underlying matrix dynamics possessing a global unitary invariance.

Trace dynamics does not specify the fundamental Lagrangian for physical interactions. Also, it does not include gravity, although it operates at the Planck scale. The theory also leaves a few important questions unanswered. What is the origin of the small anti-self-adjoint  component of the Hamiltonian at the Planck scale? Why does spontaneous localisation take place only for fermions, but not for bosons? Why should the norm of the state vector be preserved despite the presence of the 
anti-Hermitean fluctuations? In the next section, we demonstrate how to include gravity in trace dynamics, using the principles of Connes' non-commutative geometry - this leads us to a candidate quantum theory of gravity, for which we specify a Lagrangian. We also answer the open questions left unanswered by trace dynamics, as mentioned in the previous lines.

\section{Incorporating gravity in trace dynamics}
In the Introduction, we have argued that there ought to exist an equivalent reformulation of quantum (field) theory which does not refer to classical space-time. One possible way to arrive at such a reformulation is to raise space-time points to the status of non-commuting matrices/operators, in the spirit of what was done in trace dynamics above for material particles / matter fields. 
Non-commutative geometry [NCG] allows for such a possibility for space-time and its geometry. In other words, Connes' NCG
program does for space-time what trace dynamics does for matter fields. We propose to put non-commutative geometry together with trace dynamics, and propose a matrix dynamics for matter Dirac fermions and the (non-commutative) space-time geometry  produced by them. This new theory operates at Planck time/energy scales, just as TD does. The statistical thermodynamics of this new theory - i.e. coarse-graining over times larger than Planck time, provides us with a candidate quantum theory of gravity, which is also the sought for quantum (field) theory without classical time \cite{maithresh2019}.

Henceforth we will consider a Euclidean space-time, and Euclidean general relativity. The case of Lorentzian space-times remains to be dealt with. In NCG \cite{Connes2000}, the definition of a spectral action derives from the spectral definition of infinitesimal distance using the distance operator $d\hat{s}$. This operator is  related to the Dirac operator $D$ as $d\hat{s}=D^{-1}$, thus providing a definition of distance - equivalent to the standard definition of distance [in terms  of the metric] - as and when a Riemannian geometry and a manifold exists. This spectral definition of distance continues to hold also when an underlying manifold is absent, as for instance when the algebra of coordinates does not commute. 

Next, the integral $\fint T$ of a first order infinitesimal in operator $T$ is defined to be the coefficient of the logarithmic divergence in the Trace of T \cite{Connes2000}. We may visualise the integral of an operator as if it were  the sum of its eigenvalues. The spectral action relating to gravity $S$ is defined as the slash integral
\begin{equation} 
S = \fint d\hat{s}^2 = \fint D^{-2}
\end{equation}
a definition that holds whether or not an underlying spacetime manifold is present.  When a manifold is present, this spectral action can be shown to be equal to the Einstein-Hilbert action, in the following manner. The non-commutative integral $\fint d\hat{s}^2 = \fint D^{-2}$ is given by the Wodzicki residue $Res_{W}D^{-2}$, which in turn is proportional to the volume integral of the second coefficient in the heat kernel expansion of $D^{2}$. The Lichnerowicz formula relates the square of the Dirac operator to the scalar curvature, thus enabling the remarkable result \cite{Landi1999}
\begin{equation}
\fint d\hat{s}^2 = - \frac{1}{48\pi^2} \int_{M} d^4x\; \sqrt{g} \ R
\end{equation}
In connection with  the standard model of particle physics coupling to gravity, the spectral action of the gravity sector can be written as a simple function of the square of the Dirac operator, using a cut-off function $\chi(u)$ which vanishes for large $u$ (\cite{Landi1999} and references therein):
\begin{equation}
S_G[D] =  \kappa Tr [\chi (L_P^2 D^2)]
\label{spect}
\end{equation} 
The constant $\kappa$ is chosen so as to get the correct dimensions of action, and the correct numerical coefficient. 

At curvature scales smaller than Planck curvature, this action can be related to the Einstein-Hilbert action using the heat kernel expansion:
\begin{equation}
S_G[D] = L_P^{-4} f_0 \; \kappa \int_M d^4x\ \sqrt{g} + L_p^{-2} f_2 \kappa \int_M d^4x \sqrt{g} R + ...
\label{sgd}
\end{equation}
Here, $f_0$ and $f_2$ are known functions of $\chi$ and the further terms which are of higher order in $L_p^{2}$ are ignored for the present. Also, we will not consider the cosmological constant term for the purpose of the present discussion. The development of a full quantum theory of gravity must take into account all higher order corrections. The present program is a truncated approximation to such a future theory.

Let us compare and contrast the above definition of spectral action with how a trace action is defined in Adler's theory of trace dynamics. In trace dynamics it is the Lagrangian [not the action] which is made of trace of a polynomial. Thus, the way things stand, we cannot use the spectral action directly in trace dynamics to bring in gravity into matrix dynamics. We need to think of the spectral action as a Lagrangian, and we then need to integrate that Lagrangian over time, to arrive at something analogous to the action in trace dynamics. We can convert the spectral action into a quantity with dimensions of a Lagrangian, simply by multiplying it by $c/L_p$ (equivalently, dividing by Planck time). But which time parameter to integrate the Lagrangian over? The space-time coordinates have already been assumed to be non-commuting operators, (especially in the definition of the atom of space-time-matter, as below, the case that we are interested in). So it seems as if we have a Lagrangian, but we do not have a time parameter over which to integrate the Lagrangian, so as to make an action.
Fortunately, non-commutative geometry itself comes with a ready-made answer! The required time parameter is the Connes time $\tau$, which we discussed in earlier work. In NCG, according to the 
Tomita-Takesaki theorem, there is a one-parameter group of inner automorphisms of the algebra ${\cal A}$ of the non-commuting coordinates - this serves as a `god-given' (as Connes puts it) time parameter with respect to which non-commutative spaces evolve \cite{Connes2000}. This Connes time $\tau$ has no analog in the commutative case, and we employ it here to describe evolution in trace dynamics. Thus we define the action for gravity, in trace dynamics, as
\begin{equation}
S_{GTD} =\kappa \frac{c}{L_P} \int d\tau \; Tr [\chi (L_P^2 D_B^2)]
\end{equation}
Note that $S_{GTD}$ has the correct dimensions, that of action. Also, we will henceforth denote the Dirac operator as $D_B$, instead of as $D$.

Next, we would like to derive the Lagrange equations for this trace action. For this we need to figure out what the configuration variables $q$ are. In the presence of a manifold, those variables would simply be the metric. But we no longer have that possibility here. We notice though that the operator $D_B$ is like momentum, since it has dimensions of inverse length. $D_B^2$ is like kinetic energy, so its trace is a good candidate Lagrangian. Therefore, we define a new self-adjoint  bosonic operator $q_B$, having the dimension of length, and we define a velocity $dq_B/d\tau$, which is defined to be related to the Dirac operator $D_B$ by the following new relation
\begin{equation}
D_B \equiv \frac{1}{Lc}\;  \frac{dq_B}{d\tau}
\end{equation}
where $L$ is a length scale whose significance will become clear shortly. The action for gravity in trace dynamics can now be written as
\begin{equation}
S_{GTD} =\kappa \frac{c}{L_P} \int d\tau \; Tr [\chi (L_P^2 \dot{q}^2/L^2 c^2)]
\end{equation}
where the time derivative in $\dot{q}$ now indicates derivative with respect to Connes time. For the present we will work with the function $\chi(u)=u$, leaving the consideration of convergence for future work.

We would now like to incorporate matter fermions into the theory. However we do not write the standard Dirac action for fermions, add up over all the fermions, and add that action to the gravity trace action. This is because at the Planck scale, where this theory operates, we do not make a distinction between a material particle described by a fermionic operator $q_F$, and its associated gravity $q_B$. Rather, we define an `atom of space-time-matter [STM]' by a Grassmann operator $q$ such that $q=q_B + q_F$. The natural split of $q$ into its bosonic and fermionic parts is equivalent to considering the STM atom as a combination of its matter content and its gravity part. The Hilbert space of the theory is populated by many STM atoms, each with its own operator $q_i$. The operator $q_F$ of an STM atom is used to define the `fermionic' Dirac operator $D_F$:
\begin{equation}
 D_B \equiv \frac{1}{Lc}\;  \frac{dq_F}{d\tau}
 \end{equation}
$D_B$ is defined such that in the commutative limit, it becomes the standard Dirac operator on a Riemannian manifold. $D_F$ is defined such that it gives rise to the classical action for a relativistic point particle, as we will see below.
An STM atom is assumed to be described by the following  action principle in this generalised trace dynamics including gravity:
 \begin{equation}
 \frac{S}{C_0}  =  \frac{1}{2} \int \frac{d\tau}{\tau_{Pl}} \; Tr \bigg[\frac{L_P^2}{L^2c^2}\; \left(\dot{q}_B +\beta_1 \frac{L_P^2}{L^2}\dot{q}_F\right)\; \left(\dot{q}_B +\beta_2 \frac{L_P^2}{L^2} \dot{q}_F\right) \bigg]
\end{equation}
where $\beta_1$ and $\beta_2$ are  constant self-adjoint fermionic matrices. These matrices make the Lagrangian bosonic.
The only two fundamental constants are Planck length and Planck time - these scale the length scale $L$ of the STM atom, and the Connes time, respectively. $C_0$ is a constant with dimensions of action, which will be identified with Planck's constant in the emergent theory. The Lagrangian and action are not restricted to be self-adjoint.

The canonical momenta following from this Lagrangian are constant and are given by
\begin{align}
    p_B = \frac{\delta \textbf{L}}{\delta \dot{q}_B} &= \frac{a}{2}\bigg[2\dot{q}_B + \frac{L_P^2}{L^2}(\beta_1 +\beta_2)\dot{q}_F \bigg] = c_1\\ 
    p_F = \frac{\delta \textbf{L}}{\delta \dot{q}_F} &= \frac{a}{2} \frac{L_P^2}{L^2}\bigg[\dot{q}_B (\beta_1 +\beta_2)+ \frac{L_P^2}{L^2}\beta_1 \dot{q}_F \beta_2 +  \frac{L_P^2}{L^2}\beta_2 \dot{q}_F \beta_1 \bigg]=c_2
\end{align}
where $a\equiv L_P^2/L^2c^2$.
These equations can be integrated to obtain the following solutions:
\begin{align}
     \dot{q}_B &= \frac{1}{2}\bigg[c_1 -(\beta_1+\beta_2)(\beta_1-\beta_2)^{-1}\big[2c_2 -c_1(\beta_1+\beta_2) \big](\beta_2-\beta_1)^{-1} \bigg] \label{qb} \\
     \dot{q}_F &= (\beta_1-\beta_2)^{-1}\big[2c_2-c_1(\beta_1+\beta_2)\big](\beta_2-\beta_1)^{-1} \label{qf}
\end{align}
This means that the velocities $\dot{q}_B$ and $\dot{q}_F$ are constant,  and $q_B$ and $q_F$ evolve linearly in Connes time. 
The trace Hamiltonian is given by
\begin{equation}
    \textbf{H} = \text{Tr} \frac{2}{a} \bigg[(p_B\beta_1-p_F)(\beta_2-\beta_1)^{-1}(p_B\beta_2-p_F)(\beta_1-\beta_2)^{-1}
    \bigg]
\end{equation}
The Adler-Millard charge is given by
\begin{align}
     (2/a) \; \tilde{C} &= [q_B, 2\dot{q}_B +(\beta_1+\beta_2)\dot{q}_F] -\{q_F, \dot{q}_B(\beta_1 +\beta_2)+\beta_1 \dot{q}_F \beta_2+ \beta_2 \dot{q}_F \beta_1 \} \nonumber\\
      &= [q_B,2\dot{q}_B]+[q_B,(\beta_1+\beta_2)\dot{q}_F]-\{q_F,\dot{q}_B(\beta_1+\beta_2)\} \\ \nonumber &\quad \: -\{q_F,\beta_1\dot{q}_F\beta_2+\beta_2\dot{q}_F\beta_1 \} 
\end{align}
[In Eqns. (30-33) we have suppressed the factor $L_P^2/L^2$ so as to keep the expression from being complicated; it is understood that every $\beta$ in these equations comes multiplied with this factor.]
The equation for the bosonic momentum $p_B$ can be written as a modified Dirac equation with a complex eigenvalue:
\begin{equation}
\left[D_B + \frac{L_P^2}{L^2}\frac{\beta_1+\beta_2}{2}D_F\right] \psi =  \frac{1}{L} \bigg(1+ i \frac{L_P^2}{L^2}\bigg)\psi
\label{mod}
\end{equation}
Since $D_B$ is self-adjoint, the imaginary part to the eigenvalue comes only from $D_F$, and the relative magnitudes of the real and imaginary part are dictated by the structure of operator on the left. The eigenvector depends on both $q_B$ and $q_F$. This equation plays a crucial role in the subsequent discussion below. 
We note that $D_F$ will also contribute a term to the real part of the eigenvalue; let us denote this term by $\frac{L_P^2}{L^2}\theta$. In the limit $L\gg L_P$, this term is negligible. It turns out this will be the quantum limit: the imaginary part of the eigenvalue is also ignorable, and one effectively has a self-adjoint trace Hamiltonian. In the limit $L\ll L$ this term will be significant - this happens to be the classical limit: there is also a non-negligible imaginary `fast' component at the Planck scale, which gives rise to a significant anti-self-adjoint part in the Hamiltonian. It is not clear to us at present as to what role the $\theta$ term is playing in the classical limit. It appears to modify classical general relativity, but does not affect our subsequent calculation of the black hole entropy.

We have now described our trace dynamics model including gravity. If there are $N$ STM atoms in the system, the above action is written for each atom ($L$ can be different in magnitude for different atoms), and the total action is the sum of the individual actions. We call this theory non-commutative matter gravity. 

The next step is to carry out the statistical thermodynamics for a large number of atoms, and to understand the emergent quantum gravity theory, as well as the emergence of classical space-time geometry after spontaneous localisation. The overall scenario is described in the figure below, and in its accompanying caption.
\begin{figure}[!htb]
        \center{\includegraphics[width=\textwidth]
        {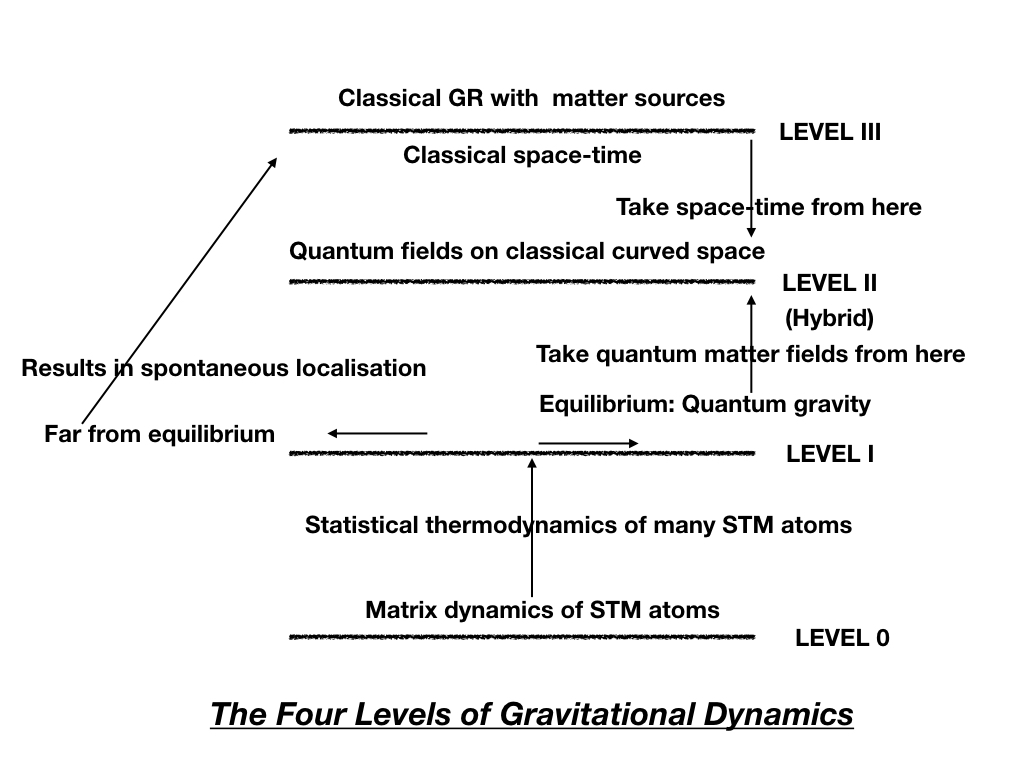}}
        \caption{\label{fig:my-label} The four levels of gravitational dynamics. In this bottom-up theory, the fundamental Level 0 describes the `classical' matrix dynamics of atoms of space-time-matter (STM). This level operates at the Planck scale. Statistical thermodynamics of these atoms brings us below Planck scale, to Level I: the emergent equilibrium theory is quantum gravity. Far from equilibrium, rapid spontaneous localisation results  in Level III: emergence of  classical space-time, obeying classical general relativity with matter sources. Level II is a hybrid level built by taking classical space-time from Level III and quantum matter fields from Level I, while neglecting the quantum gravitation of Level I. Strictly speaking, all quantum field dynamics takes place at Level I, but we approximate that to Level II.}
      \end{figure}

\section{Quantum gravity and classical general relativity from spontaneous localisation in Non-commutative Matter-Gravity}
Consider first a collection of STM atoms each of which has the property $L\gg L_P$. Then the imaginary component of the eigenvalue in the modified Dirac equation becomes negligible. As a result the trace-Hamiltonian is self-adjoint, and the Adler-Millard charge is anti-self adjoint. This is also equivalent to justifiable neglect of the "fast" imaginary component of the dynamical variables $x_r$ at the Planck scale. Hence the conditions necessary for arriving at the equilibrium statistical thermodynamics by coarse-graining over large times are satisfied.

This sets the stage for the emergence of the coarse-grained quantum gravitational dynamics at thermodynamic equilibrium. A Ward identity, which is the equivalent of the equipartition theorem, is derived. As in trace dynamics, the anti-self adjoint part of the conserved Adler-Millard charge is equipartitioned over all the degrees of freedom, and the equipartitioned value per degree of freedom is identified with Planck's constant $\hbar$. (The constant $C_0$ is now identified with $\hbar$.) At equilibrium, the standard quantum commutation relations of (an equivalent of) quantum general relativity emerge, for the canonical ensemble averages of the various degrees of freedom:
\begin{equation}
[q_B, p_B] = i \hbar; \qquad \{q_{FS}, p^f_{FAS}\} = i\hbar; \qquad \{q_{FAS}, p^f_{FS}\} = i\hbar
\end{equation}
The subscript $S/AS$ denote self-adjoint and anti-self-adjoint parts of the dynamical variables. The superscript $f$ denotes
the fermionic part of the momentum $p_F$, being the part which depends on $q_F$ but not on $q_B$: i.e. $p_F^{f} = \beta_1 \dot{q}_F\beta_2 +\beta_2 \dot{q}_F \beta_1 $.
All the other commutators and anti-commutators amongst the canonical degrees of freedom vanish at thermodynamic equilibrium. The above set of commutation relations hold for every STM atom. We note that we describe quantum general relativity in terms of these $q$ operators, and not in terms of the metric and its conjugate momenta, which are emergent concepts of Levels II and III. There is likely a possible connection between this description of quantum general relativity, and loop quantum gravity, which remains to be explored.

The mass $m$ of the STM atom is {\it defined} by $m\equiv\hbar/Lc$; and as a consequence $L$ is hence interpreted to be its Compton wavelength. Newton's gravitational constant $G$ is defined by $G\equiv L_p^2 c^3/\hbar$, and Planck mass $m_P$ by $m_P=\hbar/L_P c$. Mass and spin are both emergent concepts of Level I; at Level 0 the STM atom only has an associated length $L$ - this length is a property of both the gravity aspect and the matter aspect of the STM atom.

As a consequence of Hamilton's equations at Level 0, and as a consequence of the Ward identity mentioned above, the canonical ensemble averages of the canonical variables obey the Heisenberg equations of motion of quantum theory, these being determined by ${ H}_S$, the canonical average of the self-adjoint part of the Hamiltonian:
\begin{equation} 
i\hbar \frac{\partial q_B}{\partial \tau} = [q_B, { H}_S]; \qquad i\hbar \frac{\partial p_B}{\partial \tau} =  [p_B, { H}_S]; \qquad i\hbar \frac{\partial q_F}{\partial \tau} = [q_F,{ H}_S]; \qquad i\hbar \frac{\partial p^f_F}{\partial \tau} =  [p^f_F, { H}_S]
\end{equation}
In analogy with quantum field theory, one can transform from the above Heisenberg picture, and write a Schr\"{o}dinger equation for the wave-function $\Psi(\tau)$ of the full system:
\begin{equation}
i\hbar \; \frac{\partial \Psi}{\partial \tau} = { H_{Stot}} \Psi (\tau)
\end{equation}
where ${ H_{Stot}}$ is the sum of the self-adjoint parts of the Hamiltonians of the individual STM atoms. Since the Hamiltonian is self-adjoint the norm of the state vector is preserved during evolution.  This equation is the analog of the Wheeler-DeWitt equation in our theory, the equation being valid at thermodynamic equilibrium at Level I. This equation can possibly resolve the problem of time in quantum general relativity, because to our understanding it does not seem necessary that the physical state must be annihilated by ${ H_{Stot}}$. We have not arrived at this theory by quantising classical general relativity; rather the classical theory will emerge from here after spontaneous localisation, as we now describe.

It is known that the above emergence of quantum dynamics arises at equilibrium in the approximation that the Adler-Millard conserved charge is anti-self-adjoint, and its sef-adjoint part can be neglected. In this  approximation, the Hamiltonian is self-adjoint. Another way of saying this is that quantum dynamics arises when statistical fluctuations around equilibrium (which are governed by the self-adjoint part of $\tilde{C}$) can be neglected. These fluctuations arise when the "fast" component due to the imaginary eigenvalue in the modified Dirac equation becomes significant. This happens if $L\ll L_{Pl}$. For a single STM atom whose mass is much less than Planck mass, this would be impossible. Consider however a very large collection of STM atoms which are entangled with each other [Level I description]. The effective Compton wavelength $L_{eff}$, as it would appear in the effective modified Dirac equation,  is then given by
\begin{equation} 
\frac{1}{L_{eff}} = \sum_i \frac{1}{L_i}
\end{equation}
Clearly, if a very large number of STM atoms get entangled, their total mass can exceed Planck mass significantly; and the effective Compton wavelength becomes much smaller than Planck length. This is indicative of emergent classical behaviour, as follows. The fast varying imaginary component in the modified Dirac equation, on the Planck scale, is represented as imaginary stochastic corrections to the equilibrium quantum dynamics.

When the thermodynamical fluctuations are important, one must represent them by adding a stochastic anti-self-adjoint operator function to the total self-adjoint Hamiltonian (note that one cannot simply add the anti-self-adjoint part of the Hamiltonian to the above Schr\"{o}dinger equation, because that equation is defined for canonically averaged quantities; the only way to bring in fluctuations about equilibrium is to represent them by stochastic functions). This way of motivating spontaneous collapse is just as in trace dynamics (see Chapter 6 of \cite{Adler:04}), except that we are not restricted to the non-relativistic case, and evolution is with respect to Connes time $\tau$. Also, we do not have a classical space-time background yet; this will emerge now, as a consequence of spontaneous localisation [see also our earlier related paper `{\it Space-time from collapse of the wave function}' \cite{Singh:2019}].

Thus we can represent the inclusion of the anti-self-adjoint fluctuations in the above Schr\"{o}dinger equation by a stochastic function ${\cal H}(\tau)$ as:
\begin{equation}
i\hbar \; \frac{\partial \Psi}{\partial \tau} = [{ H_{Stot}} + {\cal H}(\tau)] \Psi (\tau)
\end{equation}
In general, this equation will not preserve norm of the state vector during evolution. However, as we noted above, every STM atom is in free particle motion. Hence it is  reasonable to demand that the state vector should preserve norm during evolution, even after the stochastic fluctuations have been added. Then, exactly as in collapse models and in trace dynamics, a new state vector is defined, by dividing $\Psi$ by its norm, so that the new state vector preserves norm. Then it follows that the new norm preserving state vector obeys an equation which gives rise to spontaneous localisation, just as in trace dynamics and collapse models (see Chapter 6 of \cite{Adler:04}). We should also mention that the gravitational origin of the anti-self-adjoint fluctuations presented here ($D_F$ is likely of gravitational origin, and relates to the anti-symmetric part of an asymmetric metric) agrees with Adler's proposal that the stochastic noise in collapse models is seeded by an imaginary component of the metric \cite{Adler2014, Adler:2018}.

It turns out to be instructive to work in the momentum basis where the state vector is labelled by the eigenvalues of the momenta $p_B$ and $p_F$. Since the Hamiltonian depends only on the momenta, the anti-self adjoint fluctuation is determined by the anti-self adjoint part of $p_F$. Hence it is reasonable to assume that spontaneous localisation takes place onto one or the other eigenvalue of $p^f_F$. No localisation takes place in $p_B$ - this helps understand the long range nature of gravity (which results from $q_B$ and the bosonic Dirac operator $D_B$). We assume that the localisation of $p^f_F$ is accompanied by the localisation of $q_F$, and hence that an emergent classical space-time 
is defined using the eigenvalues of $q_F$ as reference points.  Space-time emerges only as a consequence of the spontaneous localisation of matter fermions. Thus we are proposing that the eigenvalues of $q_F$ serve to define the space-time manifold. 

One needs to ask how a space-time manifold emerges after spontaneous localisation of fermions? Localised fermions serve as physical markers of space-time points, in the spirit of the Einstein hole argument. The recovery of the standard (commutative) Riemannian manifold is achieved because spontaneous localisation undoes the process
[${\rm Space-time\ points} \rightarrow {\rm Operators}$] achieved by going from a commutative algebra of coordinates to a non-commutative algebra. Thus, to begin with, there is a Riemannian geometry on a space-time manifold [assumed to be four-dimensional]; it is mapped to a commutative algebra, including a (diffeomorphism invariant) algebra of coordinates. When this algebra is made non-commutative, geometric concepts such as distance, metric and curvature can still be preserved, by employing the Dirac operator $D_B$. In our theory with STM atoms, each atom is by itself a non-commutative geometry, complete with these concepts. This NCG has been arrived at by raising each space-time point to operator/matrix status. What spontaneous  collapse does is to dynamically reverse this process, and restore space-time operators back to points. If sufficiently many STM atoms undergo localisation, then the manifold, metric and curvature concepts are recovered. The classical space-time manifold acts as a boundary condition which has to be fulfilled by the matrix dynamics. The space-time coordinates and metric present before the lift to the non-commutative case is restored.

 As in collapse models, the rate of spontaneous  localisation becomes significant only for objects which comprise a large number of matter fermions - hence the emergence of a classical space-time is possible only when a sufficiently macroscopic object comprising many STM atoms undergoes spontaneous localisation. The rate of localisation $T$ is in fact given by $T=
 \hbar^2/GM^3 c$ where $M$ is the total mass of the macroscopic entangled system.
  We now give a quantitative estimate as to what qualifies as sufficiently macroscopic. 

To arrive at these estimates, we recall the following two earlier equations, the action principle for the STM atom itself, and the eigenvalue equation for the full Dirac operator $D$:
\begin{equation}
\frac{L_P}{c} \frac{S}{\hbar}  =  \frac{a}{2} \int d\tau \; Tr \bigg[ \dot{q}_B^2 +\dot{q}_B\frac{L_P^2}{L^2}\beta_2\dot{q}_F +\frac{L_P^2}{L^2}\beta_1 \dot{q}_F \dot{q}_B +\frac{L_P^4}{L^4}\beta_1 \dot{q}_F \beta_2 \dot{q}_F \bigg]
\label{newacnr}
\end{equation}
\begin{equation}
[D_B + \frac{L_P^2}{L^2}\frac{\beta_1+\beta_2}{2}D_F] \psi = \lambda \psi \equiv (\lambda_R + i \lambda_I)\psi \equiv \bigg(\frac{1}{L} + i \frac{1}{L_I}\bigg)\psi
\end{equation}
In the second equation, since $D$ is bosonic, we have assumed that the eigenvalues $\lambda$ are complex numbers, and separated each eigenvalue into its real and imaginary part.  Recall that $L_I=L^3/L_P^2$. There is one such pair of equations for each STM atom, and the total action of all STM atoms will be the sum of their individual actions, with the individual action given as above.

When an STM atom undergoes spontaneous localisation, $p^f_F$ localises to a specific eigenvalue. Since $D_F$ is also made from $\dot{q}_F$, just as $p^f_F$ is, we assume that $D_F$ also localises to a specific eigenvalue, whose imaginary part is the $L_I$ introduced above. Correspondingly, the $D_B$ associated with this STM atom acquires a real eigenvalue, which we identify with the $\lambda_R\equiv 1/L$ above (setting aside for the moment the otherwise plausible situation that in general $p_F$ will also contribute to $\lambda_R$). 

The spontaneous localisation of each STM atom to a specific eigenvalue reduces the first term of the trace Lagrangian to:
\begin{equation}
Tr [\dot{q}_B^2] \rightarrow \lambda_R^2
\end{equation}
If sufficiently many STM atoms undergo spontaneous localisation to occupy the various eigenvalues $\lambda^i_R$  of the Dirac operator $D_B$, then we can conclude, from our knowledge of the spectral action in non-commutative geometry \cite{Landi1999}, that their net contribution to the trace is:
\begin{equation} 
\frac{\hbar a}{2} \;  Tr[\dot{q}_{B}^2] =\frac{\hbar }{2}  Tr [L_p^2 D_B^2] = \frac{\hbar}{2} L_p^2 \sum  (\lambda_R^i)^2 = \frac{\hbar}{2 L_p^2} \int d^4x \; \sqrt{g} \; R
\end{equation}
Thus we conclude that the Einstein-Hilbert action emerges after spontaneous localisation of the matter fermions. In that sense, gravitation is indeed an emergent phenomenon. Also, the eigenvalues of the Dirac operator $D_B$ have been proposed as dynamical observables for general relativity \cite{Rovelli}, which in our opinion is a result of great significance.
This study also demonstrates how to relate the eigenvalues of $D_B$ to the classical metric. In this sense the matrix $q_B$ captures the information of the metric field.

Let us now examine how the matter part of the general relativity action arises from the trace Lagrangian (its second and third terms) arises after spontaneous localisation. These terms are
\begin{equation}
\frac{L_p^2}{L^2}\frac{a\hbar }{2}  Tr \big[ \dot{q}_B\beta_2\dot{q}_F +\beta_1 \dot{q}_F \dot{q}_B \big] = \hbar Tr [L_p^2 \times \frac{L_P^2}{L^2}\frac{\beta_1+\beta_2}{2}D_F D_B]
\end{equation}
Spontaneous localisation sends this term to $L_p^2 \times {1}/{L_I} \times {1}/{L}$, where $L_I=L^3/L_P^2$. There will be one such term for each STM atom, and analogous to the case of $Tr D_B^2$ we anticipate that the trace over all STM atoms gives rise to the `source term'
\begin{equation}
\hbar\int \sqrt{g} \; d^4x \;\sum_i [ L_p^{-2} \times {1}/{L^i_I} \times {1}/{L^i}]
\end{equation}
Consider the term for one atom. We make the plausible assumption that spontaneous localisation localises the STM atom to a size $L_I$. This is analogous to the resolution length scale (conventionally denoted as $r_c$ in collapse models).  We know that $L_p^2 L_I = L^3$. We recall that  $L$ is the Compton wavelength $\hbar/ mc$ of the STM atom. Moreover, we propose that the classical approximation consists of replacing the inverse of the spatial volume of the localised particle - $1/L^3$, by the spatial delta function $\delta^3({\bf x} - {\bf x_0})$ so that the contribution to the matter source action becomes
\begin{equation}
\hbar \int \sqrt{g} \; d^4x \; [ L_p^{-2} \times {1}/{L_I} \times {1}/{L}] = mc \int ds
\end{equation}
which of course is the action for a relativistic point particle.

Putting everything together, we conclude that upon spontaneous localisation, the fundamental trace based action for a collection of STM atoms becomes
\begin{equation}
S = \int d^4x\; \sqrt{g} \; \bigg [\frac{c^3}{2G}R + c\; \sum_i m_i \delta^3({\bf x} - {\bf x_0})\bigg]
\end{equation}
In this way, we recover general relativity at Level III, as a result of spontaneous localisation of quantum general relativity at Level I. We should not think of the gravitational field of the STM atom as being disjoint from its related fermionic source: they both come from the same eigenvalue $\lambda$, being respectively the real and imaginary parts of this eigenvalue.

We now explain why each of the point mass localisations is a Schwarzschild black hole. The localisation takes place to a size $L_I=L^3/L_P^2$ and since the mass of the entangled atoms is much higher than Planck mass, and since $L$ is its effective Compton wavelength, $L_I$ is much smaller than Planck length. Thus we have a point mass like solution confined to below Planck length, which we have plausibly approximated by a delta function. The associated Schwarzschild radius $L_P^2/L$ is much greater than Planck length, implying that localisation happens much inside the Schwarzschild radius. The gravitational field of such a matter source is described by the emergent Einstein equations written above and is hence a Schwarzschild black hole. We note that spontaneous localisation is a process different in nature from classical gravitational collapse. Since the mass of the macroscopic object is Planck mass or higher, repeated spontaneous localisation to the same location keeps taking place at an extremely rapid rate, keeping the object as a classical black hole. In the next section we will calculate the entropy of one such black hole. On the other hand, entangled particles whose total mass is less than Planck mass, remain quantum after spontaneous localisation [i.e. do not form a black hole] because the Compton wavelength exceeds Schwarzschild radius. Thus there is a transition from classical black hole phase to quantum phase, when the net entangled mass becomes larger than Planck mass. Since there are no non-gravitational forces in our theory, spontaneous localisation of massive objects necessarily forms black holes. As and when these other interactions are included, spontaneous collapse would give rise to ordinary [non-black-hole] macroscopic objects. Interestingly, Planck length becomes the minimum observable length, because when the Compton wavelength $L$ is smaller than Planck length, the associated Schwarzschild radius exceeds Planck length, and is the observable size. At Planck mass, the Schwarzschild radius and Compton length are both Planck length  this being the minimum observable length.

We have not been able to come to a definite conclusion as regards what happens to the last term in the trace Lagrangian (\ref{newacnr}), (i.e.  $\beta_1 \dot{q}_F \beta_2 \dot{q}_F$) - after spontaneous localisation. It roughly has the structure $Tr [D_F^2]$. Adding the contribution of the eigenvalues $q_{F1}$ and $q_{F2}$ of $\beta_1 \dot{q}_F$ and $\beta_2 \dot{q}_F$ from all STM atoms, we get $Tr [q_{F1i} \; q_{F2i}]$. While we do not have a proof, we suggest that this could give rise to the cosmological constant term of general relativity. If this were to be true, then we can schematically sum up the overall picture as
\begin{equation}
S_{NMG} = \int d\tau \sum_i Tr D^2_i  \quad  {\mathbf \longrightarrow} \quad \int d\tau \bigg[ \frac{c^3}{2G}\int d^4x \; \sqrt{g} \; [R-2\Lambda] + \int d^4x \; \sqrt{g} \; L_{matter}\bigg]
\end{equation}
Here, $S_{NMG}$ on the left is the total action of all STM atoms in this Non-commutative Matter-Gravity. The action on the right side of the arrow describes classical general relativity with a cosmological constant and point matter sources, and is what emerges after spontaneous localisation.  Our theory thus elegantly unifies in a simple way, the disjoint matter - gravity descriptions on the right hand side, by bringing them together as $\sum Tr\; D_i^2$. Note that, unlike the action on the left hand side of the arrow,  the right hand side of the above equation is in no way the sum of contribution of individual STM atoms: the matter part is a sum, but the gravity part is not. Undoubtedly then, the gravity part is an emergent condensate. It simply cannot be quantised. The right hand is the [commutative] action at Level III covariant under general coordinate transformation of commuting coordinates. Whereas the left hand side action at Level I is covariant under general coordinate transformations of {\it non-commuting} coordinates. It is interesting that the transition from a non-commutative geometry to a commutative geometry is caused by spontaneous localisation, and that statistical thermodynamics plays a central role in it.

\section{A derivation of black hole entropy}
We have a precise description of the microstates of STM atoms, because we have a well-defined action principle for them, from which their matrix dynamics can be obtained. Thus, in principle, the eigenstates and eigenvalues of the Hamiltonian are known, as described above.

The mathematical structure of our theory provides evidence that these microstates can yield the correct black hole entropy. First and foremost, both quantum theory as well as classical general relativity emerge as thermodynamic approximations to the (deterministic) underlying matrix dynamics. Also, the gravitational part of the trace Lagrangian / Hamiltonian of the STM atoms is given by the sum of the squared eigenvalues of the Dirac operator, which in the presence of a Riemannian manifold equals the Einstein-Hilbert action. This same Hamiltonian occurs in the partition function (for a canonical ensemble of the STM atoms) from which the entropy is calculated. Such suggestive evidence motivates us to the following estimate of the black hole entropy, subject to a few assumptions, which we  hope to justify rigorously in future work.

A Schwarzschild black hole, say of mass $M$ much greater than Planck mass, results from the spontaneous localisation of the entangled state of a very large number of STM atoms, each of which has a mass $m_i$ much smaller than Planck mass. Entanglement is essential, so as to speed up the rate of spontaneous collapse (same effect as amplification of spontaneous collapse in collapse models). For instance, if $\psi_{A1,2,3,...N}$
and $\psi_{B1,2,3,...N}$ are two eigenstates of the  quantum gravitational system Hamiltonian at equilibrium (Level I), then the entangled state $\psi_{A1,2,3,...N} + \psi_{B1,2,3,...N}$ is also an allowed state, but because $N$ is extremely large, the superposition is rapidly lost, and one gets a macroscopic state. Because of entanglement, collapse of any one STM atom collapses them all. In particular, the Schwarzschild black hole of mass $M$ results from one such process, where all the entangled atoms collapse to the centre of the black hole, and the sum of the masses of the atoms is $M$. We recall that an STM atom is nothing but an elementary particle (fermionic) which at Level 0 is inseparable from its non-commutative geometry (bosonic). Spontaneous localisation at Level I separates the fermionic part (which collapses to the centre of the black hole) from its bosonic part, which does not collapse, thus becoming the gravitational field of the black hole at Level III. 

The entropy of the black hole results from the microstates of the STM atoms. The calculation follows the methods of standard statistical mechanics  described above to compute the entropy given by Eqn. (\ref{entro}) above, at equilibrium. As shown above, the equilibrium phase space distribution is obtained by maximising the entropy, subject to the constraints that $\rho$ is normalised, and the canonically averaged trace Hamiltonian ${\bf H}$, the averaged Adler-Millard charge $\tilde{C}$ are conserved. These constraints are implemented by introducing the Lagrange multipliers $\tilde {\tau}$ for ${\bf H}$ and the matrix $\tilde{\lambda}$ for $\tilde{C}$. Our $\tilde{\tau}$ is the same as the 
 $\tau$ in \cite{Adler:04}; we have already used up the symbol $\tau$ for Connes time.  The Lagrange multiplier $\tilde{\tau}$ has dimensions of (1/Energy), and is the analog of inverse temperature, $\beta = 1/k_B T$, in ordinary statistical mechanics, whereas $\tilde{\lambda}$ has no analog in conventional statistical mechanics. As we will justify below, we will set $\tilde{\tau}$ to the Planck energy scale; the other  Lagrange multiplier will not concern us in the present paper, and the question of their relevance for the calculation of black hole entropy is left for a future investigation.
 
 Earlier, we have computed the trace Hamiltonian for our fundamental action for the STM atom:
 \begin{equation}
    \textbf{H} = \text{Tr} \frac{2}{a} \bigg[(p_B\beta_1-p_F)(\beta_2-\beta_1)^{-1}(p_B\beta_2-p_F)(\beta_1-\beta_2)^{-1}
    \bigg]
\end{equation}
For many atoms, we simply sum over the corresponding Hamiltonians of the above form, one for each atom. It is clear that upon spontaneous localisation of a large number of STM atoms, their net trace Hamiltonian will reach the same classical limit as the trace Lagrangian; the latter limit having been shown  above. Thus we conclude the important result that, upon spontaneous localisation, 
\begin{equation}
{\bf H} \longrightarrow  \frac{1}{\tau_{Pl}} \; \int d^4x\; \sqrt{g} \; \bigg [\frac{c^4}{16\pi G}R + c^2\; \sum_i m_i \delta^3({\bf x} - {\bf x_0})\bigg]
\label{Hamcla}
\end{equation}
which will be crucial below.

The equilibrium phase space density distribution is 
\begin{align}
    \rho = Z^{-1} \exp{(-\text{Tr} \Tilde{\lambda} \Tilde{C} -\tilde\tau \textbf{H}})  \\
    Z = \int d \mu \; \exp{(-\text{Tr} \Tilde{\lambda}\Tilde{C} -\tilde\tau \textbf{H} )} \label{z}
\end{align}
 The partition function should be dimensionless; so we divide  it by a fiducial volume element $\Delta V^N$ in phase space: 
\begin{equation}
    Z = \frac{1}{\Delta V^N} \int d \mu \; \exp{(-\text{Tr} \Tilde{\lambda}\Tilde{C} -\tilde\tau \textbf{H} -\eta \textbf{N})} \label{znew}
\end{equation}
The entropy  $S_E$ is given by
\begin{equation}
   \frac{S_E}{k_B} =  \log{Z} - \text{Tr}\tilde{\lambda} \frac{\partial \log{Z}}{\partial \tilde{\lambda}} -\tilde \tau \frac{\partial \log{Z}}{\partial \tilde\tau}  \label{entropy}
\end{equation}
(We will suppress $k_B$ in subsequent expressions). We will now estimate the partition function for a large number $N$ of STM atoms which constitute the spontaneously collapsed black hole of mass $M$. As we have argued above, spontaneous localisation of the (fermionic part) of an STM atom occurs to one or the other eigenvalues of $D_B$; equivalently one or the other eigenvalues of the Hamiltonian for the atom. This allows us to express the form of $Z$, by labelling various contributing terms by the corresponding eigenvalue of the Hamiltonian. Assuming the eigenvalues to be discrete (this assumption is not compulsory; it only makes representation simpler), and ignoring for now the contribution from the Adler-Millard charge and from the trace fermion number,  the partition function for the $n$-th atom can be written as
\begin{equation}
Z_n = \sum_{i} \exp (-\tilde\tau {\bf H}_{ni})
\end{equation}
where ${\bf H}_{ni}$ is the contribution from the $i$-th eigenvalue of the $n$-th atom. 
The full partition function is the product of the partition functions of the individual STM atoms:
\begin{equation}
Z = Z_1 \times Z_2 \times.....\times Z_n =\bigg[ \sum_{i} \exp (-\tilde\tau {\bf H}_{1i})\bigg] \times
\bigg[\sum_{i} \exp (-\tilde\tau {\bf H}_{2i})\bigg] \times .....\times \bigg[\sum_{i} \exp (-\tilde\tau {\bf H}_{Ni})\bigg]
\end{equation}
This expression for the partition function can in principle now be used in Eqn. (\ref{entropy}); again retaining only the first and third terms in (\ref{entropy}), to get
\begin{equation}
S_E = \log{Z}  -\tilde \tau \frac{\partial \log{Z}}{\partial \tilde\tau}  \label{entsec}
\end{equation}
with $Z$ given as above.

It suffices to consider first the contribution to entropy from $Z_n$ of the $n$-th atom, and then multiply by the total number of atoms, to get the net entropy of the black hole. The dominant contribution comes from the term
\begin{equation} 
 -\tilde \tau \frac{\partial \log{Z_n}}{\partial \tilde\tau}  =  \tilde{\tau} \; Z^{-1}  \sum_i {\bf H}_{ni} \exp (-\tilde\tau {\bf H}_{ni})
\end{equation}

We now note that the underlying matrix dynamics of the STM atoms at Level 0 takes place at  the Planck scale. Moreover, the emergence of quantum theory at Level I takes place only on time scales much larger than Planck time. This is confirmed also in \cite{Adler:04} where in Section. 5.1 Adler writes, and we quote: "We identify the time-scale $\tau$ and mass $\tau^{-1}$ with the "fast" or "high" physical scale given by the Planck scale, and we assume that the underlying theory develops a mass  hierarchy, so that observed physics corresponds to "slow" components ....that are very slowly varying in comparison to time $\tau$". Thus, analogously, in our case too, we set $\tilde{\tau}$ = $\tau_{Pl}$. Since the ${\bf H}_{ni}$ are much smaller than $1/\tilde{\tau}_{Pl}$ the exponent above can be set to unity to a very good approximation, and we get the contribution to the entropy from the $n$-th atom to be, with $N_0$ being the number of states,
\begin{eqnarray} 
S_{En} = \tau_{Pl} N_{0}^{-1} \sum_i {\bf H}_{ni}  &=& \tau_{Pl} N_{0}^{-1} \times \frac{\hbar}{\tilde{\tau}_{Pl}} Tr{L_p^2 D_n^2}\nonumber\\ &=& \frac{1}{N_0} \int d^4x\; \sqrt{g}\bigg [\frac{c^4}{16\pi G}R + c^2\; \sum_i m_i \delta^3({\bf x} - {\bf x_0})\bigg]
\end{eqnarray} 
The net entropy is $N S_{En}$, with $N$ being the number of STM atoms, and {\it if we assume} that $N$ equals $N_0$ to a very high accuracy,  the black hole entropy is then given by
\begin{eqnarray}
S_E =   \int d^4x\; \sqrt{g} \; \bigg [\frac{c^4}{16\pi G}R + c^2\; \sum_i m_i \delta^3({\bf x} - {\bf x_0})\bigg]
\end{eqnarray} 

Next, we recall that we are working in a Euclidean 4-d space-time. Now, in this case, it is known from the literature that for a Schwarzschild black hole sourced by a point mass source at the centre, the Euclidean gravitational action is precisely equal to one-fourth the black-hole area: see Eqn. (2.7) in   \cite{Castro}
\begin{equation}
 \int d^4x\; \sqrt{g} \; \bigg [\frac{c^4}{16\pi G}R\bigg ] = \frac{\rm Area}{4L_P^2}
\end{equation}
In writing this relation it has been assumed that the Euclidean time $t_E$ has been compactified along a circle and is related to the Hawking temperature of the black hole as $2\pi t_E = \hbar /k_B T_H = 8\pi GM/c^3$.
This immediately implies that the gravitational contribution to the entropy is the same as in the 
Bekenstein-Hawking formula.

Strictly speaking, we should be able to derive the time-temperature relation in our theory, as well as prove the existence of Hawking radiation, without having to appeal to Level II physics - i.e. quantum field theory in curved space-time. In fact, we can infer Hawking radiation in our approach, on the basis of the fluctuation-dissipation theorem. The formation of a black hole of mass $M$ from a large number of entangled STM atoms, is a dissipative process (drag) which takes the atoms away from equilibrium, via spontaneous localisation. Hawking radiation (fluctuation) is the opposite of spontaneous localisation, and it takes the STM atoms back towards equilibrium, via black hole evaporation. Assuming that the black hole emits black-body radiation at a temperature $T$, we can heuristically relate $T$ to $M$ through the Einstein-Smoluchowski relation:
\begin{equation}
D = \mu k_B T
\end{equation}
with $D$ being the (quantum) diffusion constant, and the mobility $\mu$ is the ratio of the particle's terminal drift velocity to an applied force. Naively, a Newtonian estimate of the force on a test particle of unit mass would be $GM/R^2= c^4/ 4GM$ which as is known, is also the surface gravity of the black hole. Assuming that the terminal drift speed is $c$, and assuming that the diffusion constant per unit mass is $\hbar$, we get the Hawking temperature of the black hole to be $\hbar c^3/4GM$ which matches with the correct result up to a factor of $2\pi$. 

There is an instructive analogy between the fluctuation-dissipation theorem when applied to the Brownian motion of a particle in a fluid, and when applied to a black hole resulting from spontaneous collapse of STM atoms. The STM atoms are like the molecules of the fluid. The black hole is like the particle. Molecules move around in physical space, and interact via collisions. STM atoms evolve in Hilbert space and interact via entanglement. When the particle moves through the fluid, it experiences a drag, because of the dissipative frictional force caused by its collision with the molecules. A quantitative measure of the drag is the mobility $\mu$. When a highly entangled state of many STM atoms evolves unitarily in Hilbert space, it experiences a non-unitary stochastic noise (see III) which causes it to spontaneously collapse to a black hole. This non-unitary noise is the drag / dissipation whose strength (in collapse models) is measured by the collapse rate $M\gamma$, where $\gamma$ is the spontaneous collapse rate per unit mass.  In the case of the fluid, the particle, if initially at rest, experiences Brownian motion (fluctuation) because of random collisions with the molecules. Thus the dissipation and fluctuation have the same origin - collision with molecules of the fluid - and are hence related to each other. Correspondingly, in the case of matrix dynamics, the highly entangled state (i.e. the black hole), if not changing with Connes time (i.e. at rest) in Hilbert space, experiences the self-adjoint part of the stochastic noise (see III), which is then possibly the cause of black hole evaporation. Thus, both Hawking radiation and spontaneous collapse have the same origin - the stochastic (real + imaginary) noise which results from the STM atoms deviating from equilibrium - and hence the two are related.

\section{Concluding Remarks}
Spontaneous localisation was originally proposed as an ad hoc but falsifiable mechanism for explaining the puzzling absence of position superpositions in the macroscopic world, and to understand the quantum-to-classical transition. In order to overcome the phenomenological nature of the proposal, the theory of trace dynamics derives quantum field theory, as well as spontaneous localisation, as the statistical thermodynamics of an underlying classical matrix dynamics. However, trace dynamics is formulated on a classical Minkowski space-time and does not include gravity. We have exploited the spectral action in non-commutative geometry, as well as the absolute Connes time parameter, to incorporate gravity into trace dynamics. This has led us to the present quantum theory of gravity, in which spontaneous localisation is an inevitable consequence of the underlying non-commutative gravity. Black holes arise from spontaneous localisation of many entangled STM atoms, and in the present paper we have attempted to estimate black hole entropy from the microstates of the STM atoms. This has involved certain assumptions, which we hope to address rigorously in future work.

Another line of reasoning also suggests that spontaneous localisation is an inevitable consequence of quantum gravity. The absence of macroscopic position superpositions of material bodies is a pre-requisite for the existence of classical space-time (manifold as well as geometry, this being a consequence of the Einstein hole argument). If we assume space-time to be emergent from quantum gravity, then the latter must also explain why macroscopic bodies  do not exhibit position superpositions. This is a challenging task for theories of quantum gravity which are arrived at by quantizing classical gravity (unless one invokes the many-worlds interpretation). That essentially leaves spontaneous localisation as the only option for achieving a classical universe from quantum gravity. The same process which solves the quantum measurement problem, also explains the emergence of classical space-time. We can say that the quantum measurement problem is solved by quantum gravity. The subject of quantum foundations is not disconnected from quantum gravity; rather the two are directly related.

Our proposal for non-commutative matter-gravity is testable and falsifiable, for the following reasons:

1. Spontaneous localisation (the GRW theory) is a prediction of this theory, and the GRW theory is being tested in labs currently. If the GRW theory is ruled out by experiments, our  proposal will be ruled out too.

2. We have predicted the novel phenomena of quantum interference in time, and spontaneous collapse in time. This is discussed in \cite{Singh:2019}  and is falsifiable.

3. The theory predicts the Karolyhazy length as a minimum length, as a consequence of the relation between $L$ and $L_I$. This is testable and falsifiable, as discussed in \cite{Singh2019K}.

4. This theory predicts that dark energy is a quantum gravitational phenomenon, as discussed in \cite{Singh:2019DE}.

An overview of the proposed theory is available also in \cite{Simngh2019sqg} andt \cite{Singh2019qfqg}  and in the video lecture available at
\href{https://www.youtube.com/watch?v=lJk_mE8K8uw}{Spontaneous quantum gravity}

\bigskip

\bigskip

\centerline{\bf REFERENCES}

\bibliography{biblioqmtstorsion}

\def\polhk#1{\setbox0=\hbox{#1}{\ooalign{\hidewidth
  \lower1.5ex\hbox{`}\hidewidth\crcr\unhbox0}}} \def\cprime{$'$}
  \def\cprime{$'$}
\begin{thebibliography}{23}%
\makeatletter
\providecommand \@ifxundefined [1]{%
 \@ifx{#1\undefined}
}%
\providecommand \@ifnum [1]{%
 \ifnum #1\expandafter \@firstoftwo
 \else \expandafter \@secondoftwo
 \fi
}%
\providecommand \@ifx [1]{%
 \ifx #1\expandafter \@firstoftwo
 \else \expandafter \@secondoftwo
 \fi
}%
\providecommand \natexlab [1]{#1}%
\providecommand \enquote  [1]{``#1''}%
\providecommand \bibnamefont  [1]{#1}%
\providecommand \bibfnamefont [1]{#1}%
\providecommand \citenamefont [1]{#1}%
\providecommand \href@noop [0]{\@secondoftwo}%
\providecommand \href [0]{\begingroup \@sanitize@url \@href}%
\providecommand \@href[1]{\@@startlink{#1}\@@href}%
\providecommand \@@href[1]{\endgroup#1\@@endlink}%
\providecommand \@sanitize@url [0]{\catcode `\\12\catcode `\$12\catcode
  `\&12\catcode `\#12\catcode `\^12\catcode `\_12\catcode `\%12\relax}%
\providecommand \@@startlink[1]{}%
\providecommand \@@endlink[0]{}%
\providecommand \url  [0]{\begingroup\@sanitize@url \@url }%
\providecommand \@url [1]{\endgroup\@href {#1}{\urlprefix }}%
\providecommand \urlprefix  [0]{URL }%
\providecommand \Eprint [0]{\href }%
\providecommand \doibase [0]{http://dx.doi.org/}%
\providecommand \selectlanguage [0]{\@gobble}%
\providecommand \bibinfo  [0]{\@secondoftwo}%
\providecommand \bibfield  [0]{\@secondoftwo}%
\providecommand \translation [1]{[#1]}%
\providecommand \BibitemOpen [0]{}%
\providecommand \bibitemStop [0]{}%
\providecommand \bibitemNoStop [0]{.\EOS\space}%
\providecommand \EOS [0]{\spacefactor3000\relax}%
\providecommand \BibitemShut  [1]{\csname bibitem#1\endcsname}%
\let\auto@bib@innerbib\@empty
\bibitem [{\citenamefont {Ghirardi}\ \emph {et~al.}(1986)\citenamefont
  {Ghirardi}, \citenamefont {Rimini},\ and\ \citenamefont
  {Weber}}]{Ghirardi:86}%
  \BibitemOpen
  \bibfield  {author} {\bibinfo {author} {\bibfnamefont {Gian~Carlo.}\
  \bibnamefont {Ghirardi}}, \bibinfo {author} {\bibfnamefont {Alberto}\
  \bibnamefont {Rimini}}, \ and\ \bibinfo {author} {\bibfnamefont {Tullio}\
  \bibnamefont {Weber}},\ }\bibfield  {title} {\enquote {\bibinfo {title}
  {Unified dynamics for microscopic and macroscopic systems},}\ }\href@noop {}
  {\bibfield  {journal} {\bibinfo  {journal} {Phys. Rev. D}\ }\textbf {\bibinfo
  {volume} {34}},\ \bibinfo {pages} {470--491} (\bibinfo {year}
  {1986})}\BibitemShut {NoStop}%
\bibitem [{\citenamefont {Ghirardi}\ \emph {et~al.}(1990)\citenamefont
  {Ghirardi}, \citenamefont {Pearle},\ and\ \citenamefont
  {Rimini}}]{Ghirardi2:90}%
  \BibitemOpen
  \bibfield  {author} {\bibinfo {author} {\bibfnamefont {Gian~Carlo}\
  \bibnamefont {Ghirardi}}, \bibinfo {author} {\bibfnamefont {Philip}\
  \bibnamefont {Pearle}}, \ and\ \bibinfo {author} {\bibfnamefont {Alberto}\
  \bibnamefont {Rimini}},\ }\bibfield  {title} {\enquote {\bibinfo {title}
  {Markov processes in hilbert space and continuous spontaneous localization of
  systems of identical particles},}\ }\href@noop {} {\bibfield  {journal}
  {\bibinfo  {journal} {Phys. Rev. A}\ }\textbf {\bibinfo {volume} {42}},\
  \bibinfo {pages} {78--89} (\bibinfo {year} {1990})}\BibitemShut {NoStop}%
\bibitem [{\citenamefont {Pearle}(1989)}]{Pearle:89}%
  \BibitemOpen
  \bibfield  {author} {\bibinfo {author} {\bibfnamefont {Philip}\ \bibnamefont
  {Pearle}},\ }\bibfield  {title} {\enquote {\bibinfo {title} {Combining
  stochastic dynamical state-vector reduction with spontaneous localization},}\
  }\href@noop {} {\bibfield  {journal} {\bibinfo  {journal} {Phys. Rev. A}\
  }\textbf {\bibinfo {volume} {39}},\ \bibinfo {pages} {2277--2289} (\bibinfo
  {year} {1989})}\BibitemShut {NoStop}%
\bibitem [{\citenamefont {Bassi}\ and\ \citenamefont
  {Ghirardi}(2003)}]{Bassi:03}%
  \BibitemOpen
  \bibfield  {author} {\bibinfo {author} {\bibfnamefont {Angelo}\ \bibnamefont
  {Bassi}}\ and\ \bibinfo {author} {\bibfnamefont {Gian~Carlo}\ \bibnamefont
  {Ghirardi}},\ }\bibfield  {title} {\enquote {\bibinfo {title} {Dynamical
  reduction models},}\ }\href@noop {} {\bibfield  {journal} {\bibinfo
  {journal} {Phys. Rep.}\ }\textbf {\bibinfo {volume} {379}},\ \bibinfo {pages}
  {257--426} (\bibinfo {year} {2003})}\BibitemShut {NoStop}%
\bibitem [{\citenamefont {Bassi}\ \emph {et~al.}(2013)\citenamefont {Bassi},
  \citenamefont {Lochan}, \citenamefont {Satin}, \citenamefont {Singh},\ and\
  \citenamefont {Ulbricht}}]{RMP:2012}%
  \BibitemOpen
  \bibfield  {author} {\bibinfo {author} {\bibfnamefont {A.}~\bibnamefont
  {Bassi}}, \bibinfo {author} {\bibfnamefont {K.}~\bibnamefont {Lochan}},
  \bibinfo {author} {\bibfnamefont {S.}~\bibnamefont {Satin}}, \bibinfo
  {author} {\bibfnamefont {T.~P.}\ \bibnamefont {Singh}}, \ and\ \bibinfo
  {author} {\bibfnamefont {H.}~\bibnamefont {Ulbricht}},\ }\bibfield  {title}
  {\enquote {\bibinfo {title} {Models of wave function collapse, underlying
  theories, and experimental tests},}\ }\href@noop {} {\bibfield  {journal}
  {\bibinfo  {journal} {Rev. Mod. Phys.}\ }\textbf {\bibinfo {volume} {85}},\
  \bibinfo {pages} {471} (\bibinfo {year} {2013})}\BibitemShut {NoStop}%
\bibitem [{\citenamefont {Carlesso}\ and\ \citenamefont
  {Paternostro}(2019)}]{Mauro2019}%
  \BibitemOpen
  \bibfield  {author} {\bibinfo {author} {\bibfnamefont {Matteo}\ \bibnamefont
  {Carlesso}}\ and\ \bibinfo {author} {\bibfnamefont {Mauro}\ \bibnamefont
  {Paternostro}},\ }\bibfield  {title} {\enquote {\bibinfo {title}
  {Opto-mechanical tests of collapse models},}\ }\href@noop {} {\ ,\ \bibinfo
  {pages} {arXiv:1906.11041} (\bibinfo {year} {2019})}\BibitemShut {NoStop}%
\bibitem [{\citenamefont {Singh}(2015)}]{Singh:2012}%
  \BibitemOpen
  \bibfield  {author} {\bibinfo {author} {\bibfnamefont {T.~P.}\ \bibnamefont
  {Singh}},\ }\bibfield  {title} {\enquote {\bibinfo {title} {The problem of
  time and the problem of quantum measurement},}\ }in\ \href@noop {} {\emph
  {\bibinfo {booktitle} {Re-thinking time at the interface of physics and
  philosophy}}},\ \bibinfo {series and number} {(arXiv:1210.81110)},\ \bibinfo
  {editor} {edited by\ \bibinfo {editor} {\bibfnamefont {T.}~\bibnamefont
  {Filk}}\ and\ \bibinfo {editor} {\bibfnamefont {A.}~\bibnamefont {von
  Muller}}}\ (\bibinfo  {publisher} {Berlin-Heidelberg:Springer},\ \bibinfo
  {year} {2015})\BibitemShut {NoStop}%
\bibitem [{\citenamefont {Singh}(2006 [arXiv:gr-qc/0510042])}]{Singh:2006}%
  \BibitemOpen
  \bibfield  {author} {\bibinfo {author} {\bibfnamefont {T.~P.}\ \bibnamefont
  {Singh}},\ }\bibfield  {title} {\enquote {\bibinfo {title}
  {[arxiv:gr-qc/0510042]},}\ }\href@noop {} {\bibfield  {journal} {\bibinfo
  {journal} {Bulg. J. Phys.}\ }\textbf {\bibinfo {volume} {33}},\ \bibinfo
  {pages} {217} (\bibinfo {year} {2006 [arXiv:gr-qc/0510042]})}\BibitemShut
  {NoStop}%
\bibitem [{\citenamefont {Maithresh}\ and\ \citenamefont
  {Singh}(2019)}]{maithresh2019}%
  \BibitemOpen
  \bibfield  {author} {\bibinfo {author} {\bibfnamefont {Palemkota}\
  \bibnamefont {Maithresh}}\ and\ \bibinfo {author} {\bibfnamefont
  {Tejinder~P.}\ \bibnamefont {Singh}},\ }\bibfield  {title} {\enquote
  {\bibinfo {title} {Proposal for a new qantum theory of gravity iii: Equations
  for quantum gravity, and the origin of spontaneous localisation},}\
  }\href@noop {} {\ \textbf {\bibinfo {volume} {arXiv:1908.04309 Zeitschrift
  fur Naturforschung A 10.1515/zna-2019-0267}} (\bibinfo {year}
  {2019})}\BibitemShut {NoStop}%
\bibitem [{\citenamefont {Adler}(2004)}]{Adler:04}%
  \BibitemOpen
  \bibfield  {author} {\bibinfo {author} {\bibfnamefont {Stephen~L.}\
  \bibnamefont {Adler}},\ }\href@noop {} {\emph {\bibinfo {title} {Quantum
  theory as an emergent phenomenon}}}\ (\bibinfo  {publisher} {Cambridge
  University Press},\ \bibinfo {year} {2004})\BibitemShut {NoStop}%
\bibitem [{\citenamefont {Adler}(1994)}]{Adler:94}%
  \BibitemOpen
  \bibfield  {author} {\bibinfo {author} {\bibfnamefont {S.~L.}\ \bibnamefont
  {Adler}},\ }\bibfield  {title} {\enquote {\bibinfo {title} {Generalized
  quantum dynamics},}\ }\href@noop {} {\bibfield  {journal} {\bibinfo
  {journal} {Nucl. Phys. B}\ }\textbf {\bibinfo {volume} {415}},\ \bibinfo
  {pages} {195} (\bibinfo {year} {1994})}\BibitemShut {NoStop}%
\bibitem [{\citenamefont {Pearle}(2005
  [arXiv:quant-ph/0602078])}]{Pearle:2005}%
  \BibitemOpen
  \bibfield  {author} {\bibinfo {author} {\bibfnamefont {P.}~\bibnamefont
  {Pearle}},\ }\href@noop {} {\bibfield  {journal} {\bibinfo  {journal} {Stud.
  Hist. Philos. Mod. Phys.}\ }\textbf {\bibinfo {volume} {36}},\ \bibinfo
  {pages} {716} (\bibinfo {year} {2005 [arXiv:quant-ph/0602078]})}\BibitemShut
  {NoStop}%
\bibitem [{\citenamefont {Connes}(2000)}]{Connes2000}%
  \BibitemOpen
  \bibfield  {author} {\bibinfo {author} {\bibfnamefont {A.}~\bibnamefont
  {Connes}},\ }\enquote {\bibinfo {title} {Visions in mathematics - gafa 2000
  special volume, part ii},}\ \ (\bibinfo  {publisher} {Springer},\ \bibinfo
  {year} {2000})\ Chap.\ \bibinfo {chapter} {Non-commutative geometry 2000},
  pp.\ \bibinfo {pages} {481 Eds. N. Alon and J. Bourgain and A. Connes and M.
  Gromov and V. Milman, arXiv:math/0011193}\BibitemShut {NoStop}%
\bibitem [{\citenamefont {Landi}(2002 gr-qc/9906044)}]{Landi1999}%
  \BibitemOpen
  \bibfield  {author} {\bibinfo {author} {\bibfnamefont {Giovanni}\
  \bibnamefont {Landi}},\ }\bibfield  {title} {\enquote {\bibinfo {title}
  {Eigenvalues as dynamical variables},}\ }\href@noop {} {\bibfield  {journal}
  {\bibinfo  {journal} {Lect. Notes Phys.}\ }\textbf {\bibinfo {volume}
  {596}},\ \bibinfo {pages} {299} (\bibinfo {year} {2002
  gr-qc/9906044})}\BibitemShut {NoStop}%
\bibitem [{\citenamefont {Singh}(2019{\natexlab{a}})}]{Singh:2019}%
  \BibitemOpen
  \bibfield  {author} {\bibinfo {author} {\bibfnamefont {Tejinder~P.}\
  \bibnamefont {Singh}},\ }\bibfield  {title} {\enquote {\bibinfo {title}
  {Space-time from collapse of the wave-function},}\ }\href@noop {} {\bibfield
  {journal} {\bibinfo  {journal} {Zeitschrift f\"ur Naturforschung A}\ }\textbf
  {\bibinfo {volume} {74}},\ \bibinfo {pages} {147 arXiv.org:1809.03441}
  (\bibinfo {year} {2019}{\natexlab{a}})}\BibitemShut {NoStop}%
\bibitem [{\citenamefont {Adler}(arXiv:1401.0353 [gr-qc] 2014)}]{Adler2014}%
  \BibitemOpen
  \bibfield  {author} {\bibinfo {author} {\bibfnamefont {S.~L.}\ \bibnamefont
  {Adler}},\ }\bibfield  {title} {\enquote {\bibinfo {title} {Gravitation and
  the noise needed in objective reduction models},}\ }\href@noop {} {\
  (\bibinfo {year} {arXiv:1401.0353 [gr-qc] 2014})}\BibitemShut {NoStop}%
\bibitem [{\citenamefont {Adler}(2018)}]{Adler:2018}%
  \BibitemOpen
  \bibfield  {author} {\bibinfo {author} {\bibfnamefont {Stephen~L.}\
  \bibnamefont {Adler}},\ }\bibfield  {title} {\enquote {\bibinfo {title}
  {Connecting the dots: Mott for emulsions, collapse models, colored noise,
  frame dependence of measurements, evasion of the ``free will theorem''},}\
  }\href@noop {} {\bibfield  {journal} {\bibinfo  {journal} {Foundations of
  Physics}\ }\textbf {\bibinfo {volume} {48}},\ \bibinfo {pages} {1557
  arXiv:1807.11450v3} (\bibinfo {year} {2018})}\BibitemShut {NoStop}%
\bibitem [{\citenamefont {Landi}\ and\ \citenamefont
  {Rovelli}(1997)}]{Rovelli}%
  \BibitemOpen
  \bibfield  {author} {\bibinfo {author} {\bibfnamefont {Giovanni}\
  \bibnamefont {Landi}}\ and\ \bibinfo {author} {\bibfnamefont {Carlo}\
  \bibnamefont {Rovelli}},\ }\bibfield  {title} {\enquote {\bibinfo {title}
  {General relativity in terms of dirac eigenvalues},}\ }\href@noop {}
  {\bibfield  {journal} {\bibinfo  {journal} {Phys. Rev. Lett.}\ }\textbf
  {\bibinfo {volume} {78}},\ \bibinfo {pages} {3051 arXiv:gr--qc/9612034}
  (\bibinfo {year} {1997})}\BibitemShut {NoStop}%
\bibitem [{\citenamefont {Castro}(2008)}]{Castro}%
  \BibitemOpen
  \bibfield  {author} {\bibinfo {author} {\bibfnamefont {Carlos}\ \bibnamefont
  {Castro}},\ }\bibfield  {title} {\enquote {\bibinfo {title} {The euclidean
  gravitational action as black hole entropy, singularities, and space-time
  voids},}\ }\href@noop {} {\bibfield  {journal} {\bibinfo  {journal} {J. Math.
  Phys.}\ }\textbf {\bibinfo {volume} {49}},\ \bibinfo {pages} {042501 DOI:
  10.1142/S0217751X08039669} (\bibinfo {year} {2008})}\BibitemShut {NoStop}%
\bibitem [{\citenamefont {Singh}(2019{\natexlab{b}})}]{Singh2019K}%
  \BibitemOpen
  \bibfield  {author} {\bibinfo {author} {\bibfnamefont {Tejinder~P.}\
  \bibnamefont {Singh}},\ }\bibfield  {title} {\enquote {\bibinfo {title}
  {Proposal for a new qantum theory of gravity: Karolyhazy uncertainty
  relation, planck scale foam, and holography},}\ }\href@noop {} {\ \textbf
  {\bibinfo {volume} {arXiv:1910.06350 [gr-qc]}} (\bibinfo {year}
  {2019}{\natexlab{b}})}\BibitemShut {NoStop}%
\bibitem [{\citenamefont {Singh}(2019{\natexlab{c}})}]{Singh:2019DE}%
  \BibitemOpen
  \bibfield  {author} {\bibinfo {author} {\bibfnamefont {Tejinder~P.}\
  \bibnamefont {Singh}},\ }\bibfield  {title} {\enquote {\bibinfo {title} {Dark
  energy as a large scale quantum gravitational phenomenon},}\ }\href@noop {}
  {\ \textbf {\bibinfo {volume} {arXiv:1911.02955 [gr-qc]}} (\bibinfo {year}
  {2019}{\natexlab{c}})}\BibitemShut {NoStop}%
\bibitem [{\citenamefont {Singh}(2019{\natexlab{d}})}]{Simngh2019sqg}%
  \BibitemOpen
  \bibfield  {author} {\bibinfo {author} {\bibfnamefont {Tejinder~P.}\
  \bibnamefont {Singh}},\ }\bibfield  {title} {\enquote {\bibinfo {title}
  {Spontaneous quantum gravity},}\ }\href@noop {} {\ \textbf {\bibinfo {volume}
  {arXiv:1912.03266}} (\bibinfo {year} {2019}{\natexlab{d}})}\BibitemShut
  {NoStop}%
\bibitem [{\citenamefont {Singh}(2019{\natexlab{e}})}]{Singh2019qfqg}%
  \BibitemOpen
  \bibfield  {author} {\bibinfo {author} {\bibfnamefont {Tejinder~P.}\
  \bibnamefont {Singh}},\ }\bibfield  {title} {\enquote {\bibinfo {title} {From
  quantum foundations to quantum gravity},}\ }\href@noop {} {\ \textbf
  {\bibinfo {volume} {arXiv:1909.06340 [quant-ph]}} (\bibinfo {year}
  {2019}{\natexlab{e}})}\BibitemShut {NoStop}%
\end{thebibliography}%

\end{document}